\documentclass[fleqn,usenatbib]{mnras}
\pdfoutput=1
\usepackage{mathrsfs}
\usepackage[T1]{fontenc}
\usepackage{ae,aecompl}
\usepackage{graphicx}
\usepackage{amsmath}
\usepackage{amssymb}
\usepackage{array}
\usepackage[utf8]{inputenc}
\usepackage{url}
\usepackage{bm}
\usepackage{cases}
\usepackage{mathtools}
\usepackage{cuted}
\usepackage{widetext}
\usepackage{changepage}
\usepackage{soul}

\DeclareUnicodeCharacter{2212}{-}

\title[Chiappo et al.]{Dwarf spheroidal $J$-factor likelihoods for generalised NFW profiles}

\author[Chiappo et al.]{
A. Chiappo$^{1,2}$\thanks{E-mail: andrea.chiappo@fysik.su.se},
J. Cohen-Tanugi$^{3}$,
J. Conrad$^{1,2}$,
L. E. Strigari$^{4}$,
\\
$^{1}$The Oskar Klein Centre for Cosmoparticle Physics, AlbaNova, SE-106 91 Stockholm, Sweden\\
$^{2}$Department of Physics, Stockholm University, AlbaNova, SE-106 91 Stockholm, Sweden\\
$^{3}$Laboratoire Univers et Particules de Montpellier, IN2P3/CNRS et  Universit\'e de 
Montpellier, 34095 Cedex 05 Montpellier, France\\
$^{4}$Mitchell Institute for Fundamental Physics and Astronomy,  Department of Physics 
and Astronomy, Texas A\&M University,\\ College Station, TX 77845, USA
}

\date{Accepted 2019 June 24. Received 2019 June 19; in original form 2018 October 23}

\pubyear{2018}

\begin{document}
\label{firstpage}
\pagerange{\pageref{firstpage}--\pageref{lastpage}}
\maketitle

\begin{abstract}
Indirect detection strategies of particle Dark Matter (DM) in Dwarf spheroidal 
satellite galaxies (dSphs) typically entail searching for annihilation signals 
above the astrophysical background. To robustly compare model predictions with 
the observed fluxes of product particles, most analyses of astrophysical 
data -- which are generally frequentist -- rely on estimating the abundance of  
DM by calculating the so-called \textit{J-factor}. This quantity is usually 
inferred from the kinematic properties of the stellar population of a dSph 
using the Jeans equation, commonly by means of Bayesian techniques that entail 
the presence (and additional systematic uncertainty) of prior choice. Here, 
extending earlier work, we develop a scheme to derive the profile likelihood 
for $J$-factors of dwarf spheroidals for models with five or more free parameters. 
We validate our method on a publicly available simulation suite, released by 
the Gaia Challenge, finding satisfactory statistical properties for bias and 
probability coverage. We present the profile likelihood function and maximum likelihood 
estimates for the $J$-factor of ten dSphs. As an illustration, we apply these 
profile likelihoods to recently published analyses of gamma-ray data with the 
Fermi Large Area Telescope to derive new, consistent upper limits on the DM 
annihilation cross-section. We do this for a subset of systems, generally 
referred to as \textit{classical dwarfs}. The implications of these findings 
for DM searches are discussed, together with future improvements and extensions 
of this technique.
\end{abstract}

\begin{keywords}
galaxies: dwarf -- galaxies: kinematics and dynamics -- dark matter
\end{keywords}


\section{Introduction}
In recent years, the quest for astrophysical identification of Dark Matter (DM) 
has brought many groups to search for high energy photons coincident with Dwarf 
spheroidal satellite galaxies (dSphs) of the Milky Way (MW)  \citep{2000PhRvD..61b3514B,Ackermann:2015zua,Wood:2015ofa,  2015ApJ...809L...4D,Fermi-LAT:2016uux, Ahnen:2016qkx,Mora:2015vhq,Zitzer:2015eqa,Zitzer:2015uta}. These objects, in fact, 
present ideal characteristics for indirect DM detection and their $\gamma$-ray 
emission could comprise traces of an annihilating, weakly interacting massive 
particle (WIMP) DM species \citep{Bertone:2004pz, Gaskins:2016cha, Conrad:2015bsa}. 
Conventional techniques for inferring such signals entail comparing the residual 
radiation over the astrophysical background with the following term
\begin{equation}
\frac{\textrm{d}N_\gamma}{\textrm{d}E_\gamma} =
\frac{\left<\sigma v\right>}{8\pi\,m^2_\textrm{DM}}
\sum_i \textrm{B}_i 
\frac{\textrm{d}N_i}{\textrm{d}E_\gamma} J 
\quad [N_\gamma \,\textrm{cm}^{-2} \textrm{s}^{-1} \textrm{GeV}^{-1}] \, ,
\label{eq:difflux}
\end{equation}
which encodes the predicted (differential) flux of photons produced per 
annihilation event. In Eq.~\ref{eq:difflux}, $\left<\sigma v\right>$ and 
$m_\textrm{DM}$ correspond to the velocity-average annihilation cross-section 
of the DM particle and its mass; $\textrm{d}N_i/\textrm{d}E_\gamma$ represents 
the (model dependent) photon spectrum produced by the annihilation channel $i$, 
scaled by its branching ratio $\textrm{B}_i$; $J$ is the so-called \textit{J-factor}. 
This last term quantifies the amount of DM present along the \textit{line-of-sight} 
(los) and it is given by \citep{Bergstrom:2000pn}
\begin{equation}
J(\Delta\Omega, D) = \int_{\Delta\Omega} \textrm{d}\Omega 
\int_\textrm{los} \rho^2_\textrm{DM}(r(s)) \textrm{d}s \quad 
\left[\textrm{GeV}^2 \textrm{cm}^{-5} \right] \, ,
\label{eq:jfactor}
\end{equation}
where $D$ is the distance to the centre of the dSph, 
$\Delta\Omega = 2\pi\left(1-\cos \theta_\textrm{max}\right)$ defines the 
cone of observation (centred on the los) and 
$\rho_\textrm{DM}$ represents the DM density distribution within the halo containing 
a dSph. Given the indeterminacy of the latter quantity and its influence on 
$\textrm{d}N_\gamma/\textrm{d}E_\gamma$ (via Eq.~\ref{eq:difflux}), we see 
how $J$ constitutes a major source of systematic uncertainty in indirect DM searches. 

It is customary to estimate $\rho_\textrm{DM}$ from the kinematic properties of the 
stellar population of a dSph using the spherical Jeans equation, typically by means 
of a Markov chain Monte Carlo (MCMC, see for example \citealt{2015PhRvD..91h3535G} 
and \citealt{2015ApJ...808L..36B}), which results in a posterior distribution. 
Marginalisation of this probability produces a posterior of the $J$-factor which 
resembles a log-normal \citep{2015MNRAS.451.2524M}. Hence, \citet{Ackermann:2015zua} 
assumed their likelihood of $J$ to be a log-normal approximation of the 
posterior probability. Moreover, the use of Bayesian methods implies the 
introduction of prior probability densities, whose influence propagates to 
the parameters estimates. Consequentially, this 
approach enforces a specific functional form on the $J$-factor likelihood, whose 
moments are biased by the priors \citep{2015MNRAS.451.2524M}. On the other hand, 
conventional $\gamma$-rays analyses are achieved via the (prior-less) maximisation 
of the Poissonian likelihood of the expected signal photons over the astrophysical 
background -- in the case of dSphs, mainly consisting of an isotropic component and 
local point sources. Therefore, combining a frequentist $\gamma$-ray likelihood with 
the Bayesian-derived $J$-factor likelihood inevitably implies an influence of 
priors on the final results. 

An alternative, prior-free approach for constructing likelihood curves for $J$ has 
been presented by \citet[hereafter CH17]{Chiappo:2016xfs}. There, however, the 
authors considered a rigid model for the underlying DM distribution, implying a 
low-dimensionality fit of the stellar kinematic data. Since most Bayesian-based 
analyses allow for more flexible models, to be comparable, a frequentist method 
should consider an (at least) equally broad parameters space (as in 
\citealt{Geringer-Sameth:2014yza}). 

In this paper, we extend the work of CH17 improving their method by increasing 
the freedom in the model when fitting the kinematic data. Using an MCMC tool 
to sample the likelihood, we derive new profile likelihoods for $J$. 
\footnote{Given a likelihood function $L(\textrm{Data}|\alpha,\Vec{\xi})$, where $\alpha$ 
is the parameter of interest and $\Vec{\xi}$ the vector containing the nuisance 
parameter(s), the \textit{profile likelihood} is defined as 
$\Hat{L}(\textrm{Data}|\alpha) = \sup_{\Vec{\xi}}L(\textrm{Data}|\alpha,\Vec{\xi})$.}
Combining these curves with the photon likelihoods -- 
published by \citet{Ackermann:2015zua} -- we obtain new constraints on 
$\left<\sigma v\right>$. This article is organised as follows: the next 
section summarises our 
assumptions in modelling the dynamical state of dSphs and the method we devised 
to construct the $J$-factor profile likelihood; in the third section we present 
the validation of our approach on a publicly available simulation suite; section 
four describes the results of our method, including new estimates for $J$ and its 
uncertainty; in section five we combine our $J$ likelihood curves with the 
published photon data likelihoods to derive new upper limits on the DM 
annihilation cross-section; finally, in the last section we discuss the 
implications of our findings for DM searches and summarise.

\section{Method}
\label{sec2}
The modelling choices made in this project follow closely those of CH17.  
We use an unbinned likelihood, assuming a Gaussian los stellar velocity 
distribution for all dSphs, which reads
\begin{align}
\mathcal{L} (\bm{\theta})
& = -\ln L(\bm{\theta} | \mathcal{D} = (\bm{R}, \bm{v}, \bm{\epsilon})) \nonumber\\
& = \frac{1}{2} \sum^{N_\star}_{i=1} \left[
\frac{(v_i-u)^2}{\sigma^2_i} + 
\ln(2\pi\sigma^2_i) \right] \, ,
\label{eq:lnlike}
\end{align}
where $u$ is the mean of all observed los stellar velocities, $v_i$, 
in a given dSph. In Eq.~\ref{eq:lnlike}, $\bm \theta$ represents the vector 
containing the parameters of interested, while $\mathcal{D}$ is the data matrix. 
The expected velocity dispersion is expressed as the squared 
sum of the velocity measurement uncertainty, $\epsilon_i$, and the 
intrinsic los dispersion, 
$\sigma^2_\textrm{los}(R_i)$: $\sigma^2_i = 
\epsilon^2_i+\sigma^2_los(R_i)$. The 
latter term is a function of the projected radial distance $R_i$ of the star 
from the dSph's centre. 
We follow the standard procedure of deriving 
$\sigma^2_los(R_i)$ via a spherical Jeans analysis 
\citep{1987gady.book.....B}, 
which gives a rather cumbersome expression for this quantity \citep{Bonnivard:2014kza}. 
\citet{Mamon:2004xk} showed that $\sigma^2_los(R_i)$ can be cast in a 
more compact form, which reads
\begin{equation}
\sigma^2_\textrm{los}(R) = 
\frac{2G}{\Sigma(R)} \int^\infty_R 
K\left(\frac{r}{R},\frac{r_a}{R}\right)\nu(r)M(r)\frac{\textrm{d}r}{r} \, .
\label{eq:MLJeans}
\end{equation}
In Eq.~\ref{eq:MLJeans}, $K(u,w)$ is a kernel function which 
encodes information on the anisotropy of the stellar velocities. 
In this project we consider three possible scenarios: isotropic 
velocity distribution (ISO); constant degree of anisotropy $\beta$ across 
the entire dSph (CB); the Osipkov-Merritt (OM) radially increasing 
velocity anisotropy profile \citep{1979SvAL....5...42O,1985AJ.....90.1027M}. 
The corresponding functional expressions of $K$ are listed below
\begin{widetext}
\begin{subnumcases}{\label{eq:ker} K(u,w) =}
\sqrt{1-\frac{1}{u^2}} & \text{(Isotropic)} \, , \label{eq:keriso} \\
\frac{w^2+1/2}{\left (w^2+1\right)^{3/2}}\, 
\left(\frac{u^2+w^2}{u}\right)\, 
\tan^{-1} \left(\sqrt{\frac{u^2-1}{w^2+1}}\right) \, -
\,\frac{1/2}{w^2+1}\,\sqrt{1-1/ u^2} & \text{(Osipkov-Merritt)} \, , \label{eq:kerom} \\
\frac{\sqrt{1-1/u^2}}{1-2\beta} + \frac{u^{2\beta-1}}{2} 
\frac{\sqrt{\pi}\,\Gamma\left(\beta-1/2\right)}{\Gamma\left(\beta\right)}
\left(3/2-\beta\right) \,
\mathcal{I}\left(1-\frac{1}{u^2}, \frac{1}{2},\beta+\frac{1}{2}\right)
& \text{(constant-$\beta$)} \, , \label{eq:kerbeta}
\end{subnumcases}
\end{widetext}
where $\mathcal{I}$ appearing in Eq.~{\ref{eq:kerbeta}} above is the 
\texttt{Incomplete Beta function}, as derived by \citet{2006MNRAS.370.1582M}. \\
The surface number density of the system, $\Sigma(R)$, is usually derived from 
a Plummer profile \citep{1911MNRAS..71..460P}, which is given by
\begin{equation}
I(R) = \frac{L}{\pi r^2_\star} \frac{1}{\left(1+R^2/r^2_\star\right)^2} \, ,
\label{eq:surfbright}
\end{equation}
where the scale radius $r_\star$ results from fits to the photometric 
data (see \citealt{2012AJ....144....4M} and references therein for 
more information on observational features of dSphs). In turn, the 
stellar density profile $\nu(r)$ is obtained via an inverse Abel transform 
\footnote{The deprojection of a quantity $F(R)$ via inverse Abel transform 
gives $f(r) = 
- \frac{1}{\pi} \int_r^\infty \frac{\textrm{d}F/\textrm{d}R}{\sqrt{R^2-r^2}} \textrm{d}R$ 
\citep{1982MNRAS.200..361B}.} of 
Eq.~\ref{eq:surfbright}, giving
\begin{equation}
\nu(r) = \frac{3 L}{4 \pi r^3_\star} 
\frac{1}{\left(1+r^2/r^2_\star\right)^{5/2}} \, .
\label{eq:stelprof}
\end{equation}
From the term $L/I(R)$ entering Eq.~\ref{eq:MLJeans}, we see that $L$ 
has no net effect on this formula and thus can be neglected. We note that 
the use of the Plummer profile (Eq.~\ref{eq:surfbright}) can lead to 
underestimating the uncertainties on $J$ and potentially biasing its value.  
This issue originates in the additional uncertainty in modelling the surface 
density profile of dSphs, whose outer envelopes are difficult to measure 
\citep{2006A&A...459..423B,2016ApJ...817...84K}. 
For the mass $M$ of the system, one should in principle account for all massive 
components of a dSph, including stars, DM, and diffuse gas. However, 
is has been shown \citep{2011ApJ...733...46S, Battaglia:2013wqa} that 
dSphs are generally DM-dominated systems. Therefore, $M$ can be safely 
approximated with
\begin{equation}
M(r) = 4\pi\int^r_0 \rho_\textrm{DM}(r)r^2 \textrm{d}r \, .
\end{equation}
In this work we parameterise $\rho_\textrm{DM}$ with a generalised 
NFW, which reads \citep{Zhao:1995cp} 
\begin{equation}
\rho_\textrm{\tiny DM}(r) = \frac{\rho_0}{\left(r/r_0\right)^c
\left(1+\left(r/r_0\right)^a\right)^{\frac{b-c}{a}}} \, ,
\label{eq:genNFW}
\end{equation}
where $\rho_0$ is the scale density within the radius $r_0$, while $a$ 
controls the transition between the steepness of the inner part of 
the profile, $c$, and the outer one, $b$. 
Eq.~\ref{eq:genNFW} can also 
describe the stellar distribution, in which case the parameters refer 
to the visible counterpart of a dSph; Eq.~\ref{eq:stelprof} is recovered 
replacing "0" with "$\star$" and by setting $(a,b,c) = (2,5,0)$.

Given the dependence of $\sigma^2_los$ on $\rho_\textrm{DM}$, 
Eq.~\ref{eq:MLJeans} provides the link between the kinematics of 
the stellar population and the underlying DM profile. Typically, 
evaluating $J$ follows the fit of the $\rho_\textrm{DM}$ parameters 
via maximisation of $L$ (Eq.~\ref{eq:lnlike}) in a Bayesian 
framework. However, the two need not be separate operations. As shown 
by CH17, a direct likelihood treatment of $J$ is possible, provided that 
the DM particle does not interact. Such approach hinges on the observation 
that most conventional DM profiles are of the form
\begin{equation}
\rho_\textrm{\tiny DM} = 
\rho_0 f\left(x=\frac{r}{r_0}, \bm{\vartheta}\right) \, ,
\end{equation}
for some function $f$ and with $\bm{\vartheta}$ the subset of parameters in 
$\bm{\theta}$ describing $\rho_\textrm{\tiny DM}$. For example, the generalised 
NFW profile defined above (Eq.~\ref{eq:genNFW}) is given by 
$f(x;\bm{\vartheta}) = x^{-c}(1+x^a)^{(c-b)/a}$, with 
$\bm{\vartheta} = (a, b, c, r_0)$. We now see that $\rho_0$ entering 
Eq.~\ref{eq:jfactor} can be expressed as
\begin{equation}
\rho_0 = \sqrt{\frac{J}{j(D, \theta_\textrm{max}; \bm{\vartheta})}} \, .
\label{eq:rho0}
\end{equation}
where $j$ is given by 
\begin{equation}
j(D,\theta_\textrm{max}; \bm{\vartheta}) = 
2\pi \, r_0 \int_{\cos\theta_\textrm{max}}^1 \textrm{d}\cos\theta 
\int_{x_\textrm{min}}^{x_\textrm{max}} \textrm{d}x\, f^2\left(x;\bm{\vartheta}\right)
\end{equation}
with $x_\text{max/min} = 
(D/r_0)\cos\theta \pm \sqrt{(r_t/r_0)^2-(D/r_0)^2\sin^2\theta}$; $r_t$ is the cutoff radius of the system, usually assumed to 
the the tidal radius \citep{2008gady.book.....B}. \\
Inserting Eq.~\ref{eq:rho0} in Eq.~\ref{eq:MLJeans}, we explicitate 
the dependence of $\mathcal{L}$ on $J$ when evaluating Eq.~\ref{eq:lnlike}, 
along with the parameters of $\rho_\textrm{DM}$ and $K$. This expedient 
allows us to implement the profiling scheme introduced by CH17 
(hereafter \textbf{manual-profiling}), which we briefly summarise below
\begin{itemize}
\item vary $J$ over a likely range
\item determine $\mathcal{L}_\textrm{MLE}(\mathcal{J}_n|\bm{\Theta})$ for each fixed $J_\textrm{n}$ 
\item interpolate between the pairs $(J_n, \mathcal{L}_\mathrm{MLE}(J_n|\bm{\Theta}))$
\end{itemize}
where $\bm{\Theta}$ represents the nuisance parameters array. The 
final point results in the profile likelihood curve of $J$, with 
which the statistical inference can be performed \citep{Conrad:2014nna}. 
Specifically, the maximum likelihood estimate (MLE) of $J$ and its 
uncertainty can be determined. 

Differently from CH17, we do not impose specific values for the DM 
profile shape parameters $(a,b,c)$, which are left free to vary in 
the likelihood optimisation process. Undesirably, most stellar 
kinematic samples impede a robust characterisation of the DM profile 
shape. This limitation originates in the scarcity of observations, 
in particular at large (projected) radial distances from the dSph 
centre. We have observed on simulations that the paucity of stars 
in the outer regions of a dSph prevents a robust determination of 
$(a,b,c)$ in Eq.~\ref{eq:genNFW}. For data-sets with $N_\star \gtrsim 1000$, instead, the shape parameters can be constrained sufficiently well. 
A consequence of this indeterminacy is the pronounced flatness of 
the likelihood in the corresponding directions of parameters space. 
In particular, we find $a,b$ to be strongly unconstrained. 
This feature of the likelihood implies great difficulties for a 
gradient-descent-based minimiser in optimising $\mathcal{L}$. 
This inconvenience has usually been addressed using an MCMC tool, 
which, however, introduces prior-dependence on the estimates. 
Moreover, priors used in the literature have been typically derived 
from N-body simulations \citep{2008MNRAS.391.1685S,2009Sci...325..970K}. 

In this project, we attempt a more agnostic approach, where we exploit 
the ability of the MCMC to explore highly covariant, large parameter 
spaces, while producing results independent from priors. Although a 
Bayesian MCMC produces credible intervals of each parameter directly 
from the posterior distribution, without an assumption of Gaussianity 
of the likelihood and independently of the number of free parameters, 
the marginal distributions depend on the choice of priors. We achieve a 
priors-independent scan of the log-likelihood parameter space by implementing 
the \textbf{emcee} package by \citet{2013PASP..125..306F}. When sampling the
log-posterior, this MCMC engine outputs also the log-likelihood evaluations 
of the examined points. Armed with this tool, we can perform the likelihood 
optimisation in the manual-profiling scheme introduced above, for $J$ fixed. 
Moreover, we can complement that approach with an alternative one where $J$ is 
free to vary (hereafter $\bm{J}$-\textbf{sampling}), schematically described below
\begin{itemize}
\item sample the parameter space $(J,\bm{\Theta})$ of $\mathcal{L}$ with an MCMC
\item retain the likelihood evaluations of the sampled points
\item construct the lower envelope of the samples in $J$
\end{itemize}
The last point, the $J$-envelope, is obtained by ordering in $J$ 
the multidimensional ensemble of likelihood evaluations (resulting from the 
first step). Then, starting from the smallest probed $J$ and retaining 
the successive, lowermost estimates of $\mathcal{L}$, provides a curve which 
maps the trough of the likelihood along this dimension – within the sampling 
uncertainty (due to limited number of MCMC iterations). Thus, the last step of 
the list above results in a proxy for the profile likelihood of $J$ and the nuisance 
parameters array $\bm{\Theta}$, given the stellar data. This curve is equivalent 
to the profile likelihood described above and is equally suitable for performing 
parameter inference. Analogously to CH17, we reformulate the analysis by fitting 
$\mathcal{J} = \log_{10}\left(J/\left[\textrm{GeV}^2\,\textrm{cm}^{-5}\right)\right]$. 
This choice is motivated by the order of magnitude of $J$ for 
various astrophysical systems, as suggested by previous analyses: 
roughly ranging in $10^{15}$ -- $10^{21}$ $\textrm{GeV}^2 \textrm{cm}^{-5}$ 
(cf \citealt{Charles:2016pgz} or \citealt{Conrad:2015bsa}). 
\begin{figure}
\centering
\includegraphics[width=0.5\textwidth, keepaspectratio]{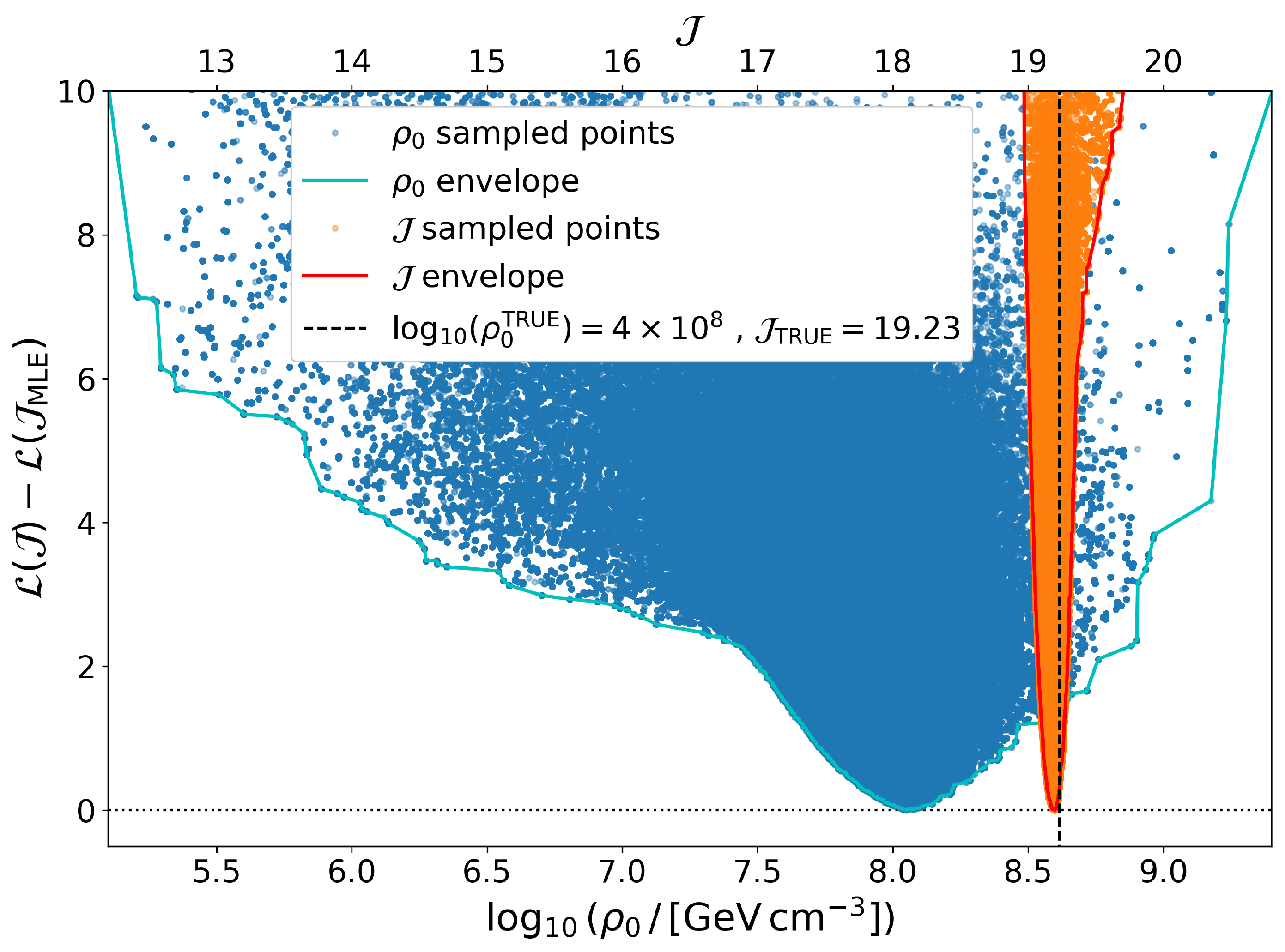}
\caption{Comparison between fitting $\rho_0$ or $\mathcal{J}$ -- together with 
the other parameters in Eq.~\ref{eq:genNFW} -- in the analysis of a mock stellar 
kinematic data-set. The blue (orange) points represent the likelihood evaluations 
projected onto the $\rho_0$ ($\mathcal{J}$) direction of the parameter space. The 
cyan (red) curve indicates the envelope of the sampled points along $\rho_0$ 
($\mathcal{J})$. The vertical dashed line corresponds to the true value of $\rho_0$ 
and correspondingly $\mathcal{J}$.}
\label{fig:fitrho0}
\end{figure}

\begin{figure}
\centering
\includegraphics[width=0.5\textwidth,keepaspectratio]{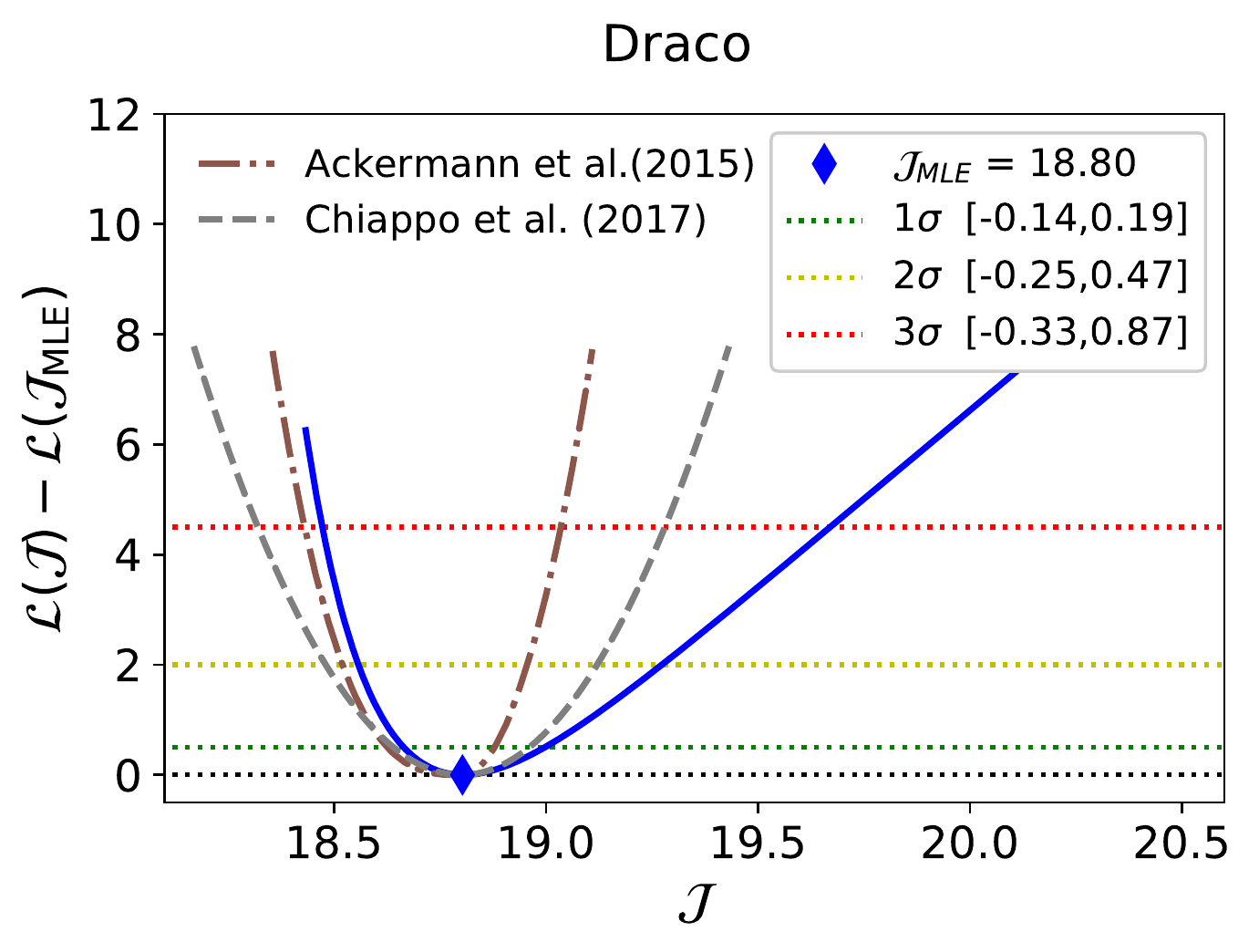}
\caption{Profile likelihood of $\mathcal{J}$ for Draco dSph. The solid 
blue curve is obtained by approximating the result of the $J$-sampling 
technique (the envelope) with Eq.~\ref{eq:linex}. We analyse the kinematic  
data assuming isotropic stellar velocities, with spatial distribution 
following a Plummer profile. For the DM component, a generalised NFW 
is adopted, all parameters of which are free in the optimisation (see 
text). For reference, the analogous result obtained by CH17 is also 
shown (dashed gray line), together with the curve used by 
\citet{Ackermann:2015zua} (dot-dashed brown line).}
\label{fig:dra}
\end{figure}

The advantage of performing an inference on $\mathcal{J}$, rather than $\rho_0$, is 
manifest in Fig.~\ref{fig:fitrho0}. This figure displays the likelihood evaluations (Eq.~\ref{eq:lnlike}) obtained leaving either $\rho_0$ (blue points) or $\mathcal{J}$ 
(orange points) free -- along with the other parameters of Eq.~\ref{eq:genNFW} -- in 
the analysis of a (simulated) stellar kinematic data-set. We note how the envelope 
of the former points (cyan curve) is much broader than the envelope of the latter 
(red curve). This difference shows how the reparameterisation in Eq.~\ref{eq:rho0} 
considerably mitigates the degeneracy between the DM normalisation parameter and 
the nuisance parameters. Moreover, we also observe how the likelihood sampling 
with respect to $\rho_0$ seems to be biased towards small values of this quantity. 
Importantly, we note that for finite MCMC sampling steps, the $J$-envelope of the 
likelihood evaluations will inevitably be a non-smooth curve. To remedy this feature, 
and contextually optimise the statistical bias and coverage, we parameterise the 
envelope of the likelihood with
\begin{equation}
h(x;p,q,r) = 
e^{-p\, x} + q\, x + r \, ,
\label{eq:linex}
\end{equation}
where $p,q,r$ are free parameters and $x = \mathcal{J} - \mathcal{J}_\textrm{MLE}$. 
The equation above represents an adaptation of the \textit{Linex loss function} 
\citep{Chang:2007abcd} and its application on Draco 
is shown in Fig.~\ref{fig:dra} (solid blue line). Comparing this curve 
with Fig. 1 of CH17 (dashed grey line), we see how varying 
even the DM profile shape parameters in the fit leads to a broadening 
of the profile likelihood. Additionally, we include the likelihood 
adopted by \citet[dot-dashed brown line]{Ackermann:2015zua}. We 
emphasise that this curve was obtained by fitting a log-normal to 
a posterior probability sampled with the Multi-level Bayesian modelling 
(MLM) proposed by \citet{2015MNRAS.451.2524M}. This aspect highlights the 
importance of determining $J$-factor likelihoods in a consistent 
manner. To determine the robustness of a this method, we assess its 
statistical performance on a simulation suite. The details of the tests 
are presented in the next section.

\section{Validation on Gaia simulations}
\label{sec3}
We validate our method on the publicly available simulation suite 
released by \textit{Gaia Challenge} \citep[hereafter GAIASIM]{2011ApJ...742...20W}
\footnote{\url{http://astrowiki.ph.surrey.ac.uk/dokuwiki/doku.php?id=workshop}}. 
The manual-profiling approach was already tested by CH17 using the same 
mock data. Since those authors considered a more restrictive (simplified) 
model, the expectation for the current, more general study is that the 
same tests would lead to larger bias and higher statistical coverage. These 
predictions are motivated by the freedom in the likelihood, on one side, 
and by its flatness in $J$ due to the indeterminacy of some nuisance 
parameters, on the other.
\begin{table*}
\caption{Models tested with the $J$-sampling MLE scheme. For each, we 
optimise the likelihood in Eq.~\ref{eq:lnlike} using data-sets containing 
$N_\star =100, 200, 500, 1000$. 
All models assume $r_0$ = 1 kpc. The entries in the fifth column ($c_\star$) 
refer to the inner slope of the Hernquist profile (Eq.~\ref{eq:genNFW}) 
describing the stellar distribution.}
\label{tab:gaia}
\setlength{\extrarowheight}{2pt}
\begin{tabular}{ r c c c c c }
\hline
Mock dSph data-sets & $\rho_0$ ($\text{M}_\odot \text{kpc}^{-3}$) & 
$\mathcal{J}_\textrm{TRUE}$  
& $r_a$ (kpc) & $c_\star$ & $r_\star$ (kpc)\\
\hline
OM Core non-Plummer 		& 4$\times 10^8$ 	& 19.23 & 0.25 		& 1 	& 0.25 	\\
OM Core Plummer-like 		& 4$\times 10^8$ 	& 19.23 & 0.25 		& 0.1 	& 0.25 	\\
Isotropic Core non-Plummer 	& 4$\times 10^8$ 	& 19.23 & $\infty$ 	& 1 	& 1 	\\
Isotropic Core Plummer-like & 4$\times 10^8$ 	& 19.23 & $\infty$ 	& 0.1 	& 1 	\\
OM Cusp non-Plummer 		& 6.4$\times 10^7$ 	& 18.83 & 0.1 		& 1 	& 0.1 	\\
OM Cusp Plummer-like 		& 6.4$\times 10^7$ 	& 18.83 & 0.1 		& 0.1 	& 0.1 	\\
Isotropic Cusp non-Plummer 	& 6.4$\times 10^7$ 	& 18.83 & $\infty$ 	& 1 	& 0.25 	\\
Isotropic Cusp Plummer-like & 6.4$\times 10^7$ 	& 18.83 & $\infty$ 	& 0.1 	& 0.25 	\\
\hline
\end{tabular}
\end{table*}
Here we assess the $J$-sampling approach by repeating the tests 
performed by CH17 on the eight models listed in Table \ref{tab:gaia}.
In line with the modelling assumptions described above (Sec.~\ref{sec2}), 
we consider only the single-component models 
released by GAIASIM and analysed by CH17. 
When fitting each mock data-set, we assume the true model in 
Eq.~\ref{eq:MLJeans}, with the exception of $\rho_\textrm{DM}$. 
For the DM profile, a generalised NFW (Eq.~\ref{eq:genNFW}) is 
adopted, but its shape parameters are free to vary. This means 
that the MCMC tool samples a total of 5 (6) parameters, namely 
$\mathcal{J}, r_0, a, b, c$ (plus $r_a$), 
whose allowed ranges are listed in Table \ref{tab:ranges}. 
\newcolumntype{C}[1]{>{\centering\arraybackslash}p{#1}}
\newcolumntype{L}[1]{>{\raggedright\arraybackslash}p{#1}}
\begin{table}
\centering
\caption{Parameters ranges allowed in all fits performed in this project.}
\label{tab:ranges}
\setlength{\tabcolsep}{2pt}
\begin{tabular}{c c L{4cm}}
$\mathcal{J}$ 	                    & $\in$ & $[10,30]$	    \\
$\log_{10} (r_0/1\, \textrm{kpc})$  & $\in$ & $[-3,2]$  	\\
$\log_{10} (r_a/1\, \textrm{kpc})$	& $\in$ & $[-3,2]$      \\ 
$\beta$                             & $\in$ & $[-9, 0.9]$   \\
$a$ 			                    & $\in$ & $(0,8]$ 		\\
$b$ 			                    & $\in$ & $[0.5,10]$ 	\\
$c$ 			                    & $\in$ & $[0,1.5)$ 	\\
\end{tabular}
\end{table}

In order to validate the statistical properties of the method, 
we estimate its bias and the statistical coverage of the 1,2,3$\sigma$ intervals. 
\footnote{Under the assumption that Wilks theorem applies \citep{Wilks:1938dza}, 
the 1,2,3$\sigma$ confidence intervals of the MLE are obtained from the 
log-likelihood ratio of the profile likelihood at the nominal values of 0.5, 2, 4.5, 
respectively.} Similarly to CH17, we do this for samples of different sizes, 
obtained by partitioning the full data-set -- the one containing 
$10^4$ stars -- into non-overlapping subsets of 100,200,500,1000 stars. 
For instance, this operation produces $N_\textrm{PE}$ = 50 (20) mock samples 
or pseudo-experiments (PE) containing $N_\star$ = 200 (500) stars.
\begin{figure*}
\includegraphics[width=\textwidth,height=\textheight,keepaspectratio]{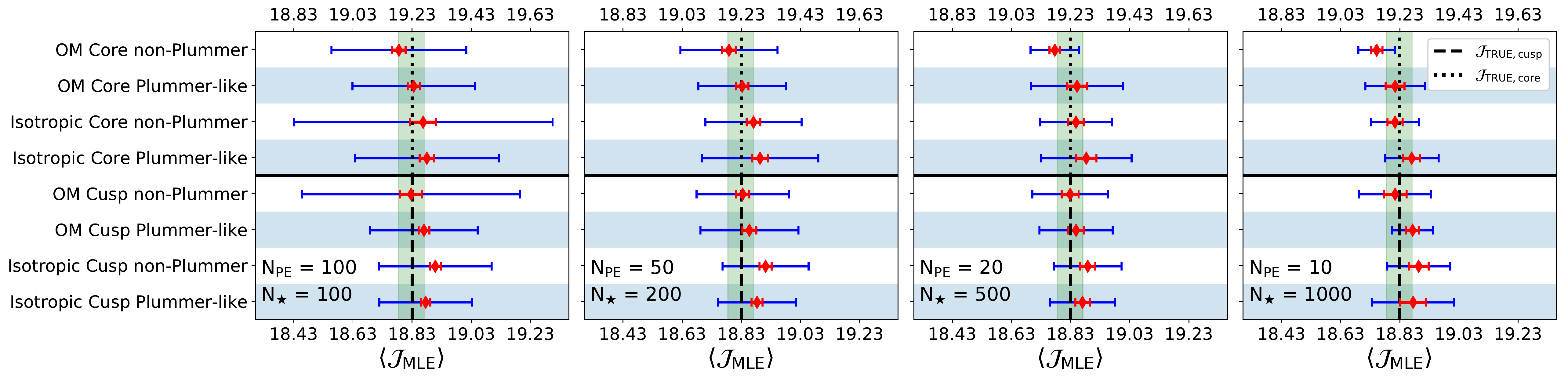}
\caption{Bias estimates of the frequentist fitting method implemented via the 
$J$-sampling scheme. In every panel, the red points represent the mean 
of the $\mathcal{J}_\textrm{MLE}$ estimates, obtained from the approximation 
with Eq.\ref{eq:linex} of the likelihood resulting from the analysis of 
each PE. The blue (red) error bars correspond to the 
root mean square (uncertainty on the mean) of the MLE values, while the 
vertical dashed (dotted) line indicates the true $\mathcal{J}$ for the Cusp 
(Core) models (see Table \ref{tab:gaia}). The vertical green band gives 
the $\mathcal{J}$ range where the bias in $J$ is $\leq$ 10\%.}
\label{fig:bias}
\end{figure*}
\begin{figure*}
\includegraphics[width=\textwidth,height=\textheight,keepaspectratio]{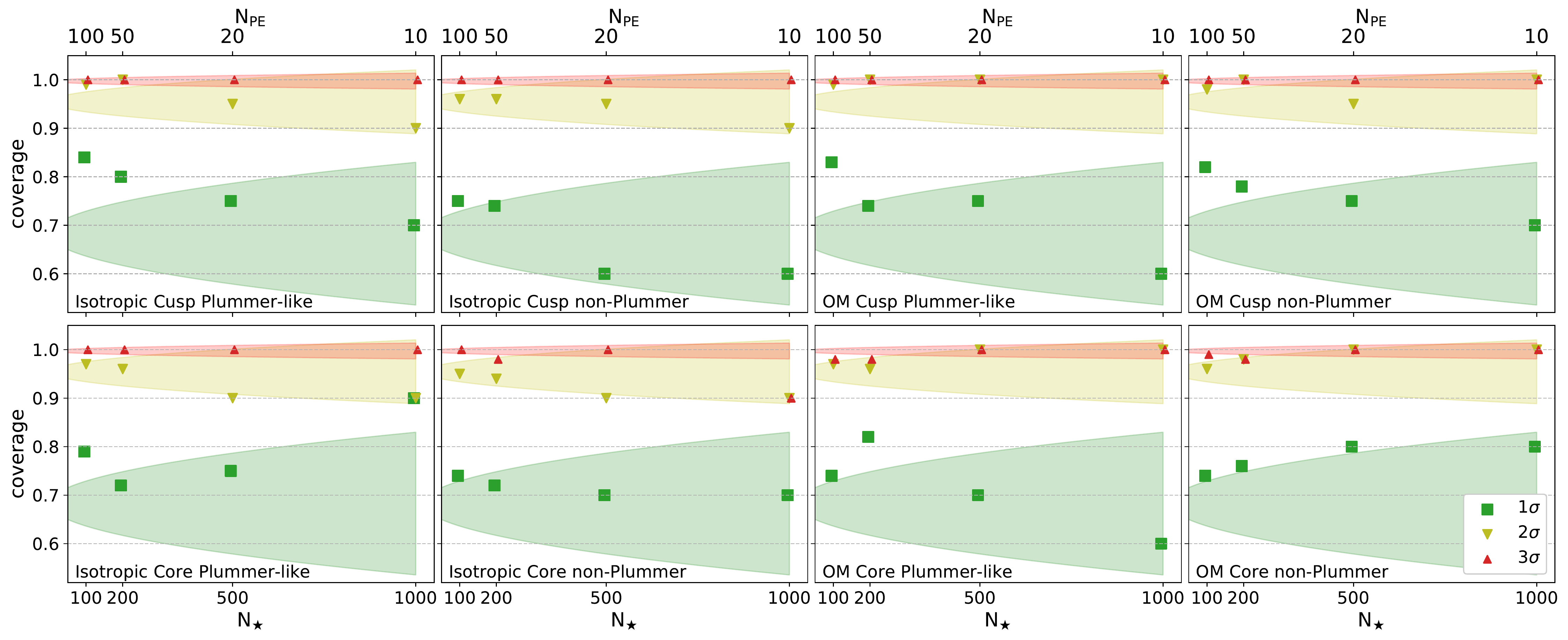}
\caption{Coverage probability as a function of the sample size. In 
every panel, the green squares, the yellow downward triangles and 
the red upward triangles indicate, respectively, the statistical 
coverage of the 1,2,3$\sigma$ intervals. The coloured bands 
represent the ranges of expected statistical coverage of an ideal 
experiment, corresponding to each $\sigma$ level. This means 
that the green, yellow and red areas are centred, respectively, 
on 68\%, 95\% and 99\%, and widen with decreasing $N_\textrm{PE}$ (see text). 
Every panel in this figure refers to one of the eight Gaia models 
considered here (see Table \ref{tab:gaia}), as indicated in 
the bottom-left corner of each plot.}
\label{fig:cov}
\end{figure*}
The outcome of these tests is shown in Figs.~\ref{fig:bias} and 
\ref{fig:cov}, displaying the bias and statistical coverage, respectively. 
All results reported in these figures were obtained by approximating 
the $\mathcal{J}$-envelope with Eq.~\ref{eq:linex}. The statistical 
coverage is generally high or within the expected range (coloured bands 
\footnote{The semi-width of the range of expected statistical coverage 
$p$ over $N$ experiments is given by $\sqrt{p(1-p)/N}$} in Fig.~\ref{fig:cov}). 
The occurrence of marginal under-coverage (at the 3$\sigma$ level in the 
Isotropic Core non-Plummer model) is plausibly a symptom of the low 
statistics regime ($N_\textrm{PE} = 10$). Importantly, we observe an 
increase of the bias for the $N_\star = 100$ and 1000 cases, as compared 
to the analogous result by CH17 (their Fig.~5). This feature is likely a 
reflection of the indeterminacy of some parameters mentioned above. 
Moreover, we recall that the MLE of a quantity becomes an unbiased estimator 
only asymptotically \citep{James:2006zz}. Whereas the uncertainty on 
$\left<\mathcal{J}_\mathrm{MLE}\right>$ is generally small (red error bars 
in Fig.~\ref{fig:bias}), we note a large scatter in the estimates (blue 
error bars in the same figure), which reduces with growing $N_\star$. 
This feature is consistent with the expectations from the validation of a 
frequentist approach. Finally, it should be remembered, as noted by CH17, 
that the GAIASIM simulations  were not generated from a Gaussian velocity 
distribution. Therefore, fitting the mock kinematic data with a Gaussian 
likelihood entails a model systematics, which could be potentially 
responsible for the observed bias.

Altogether, the results reported in this section indicate that 
the method possesses satisfactory statistical properties 
for the $N_\star$ range considered. We verified using the 
same mock data that the both bias and statistical coverage increase 
significantly when analysing smaller data-sets. This 
aspect supports the suitability of this procedure only  
for large kinematic samples. This conclusion 
necessarily imposes restrictions for the application on 
real stellar data, as detailed in the next section.

\section{Results on real kinematic data}
\label{sec4}
\begin{figure*}
\centering
\includegraphics[width=\textwidth,keepaspectratio]{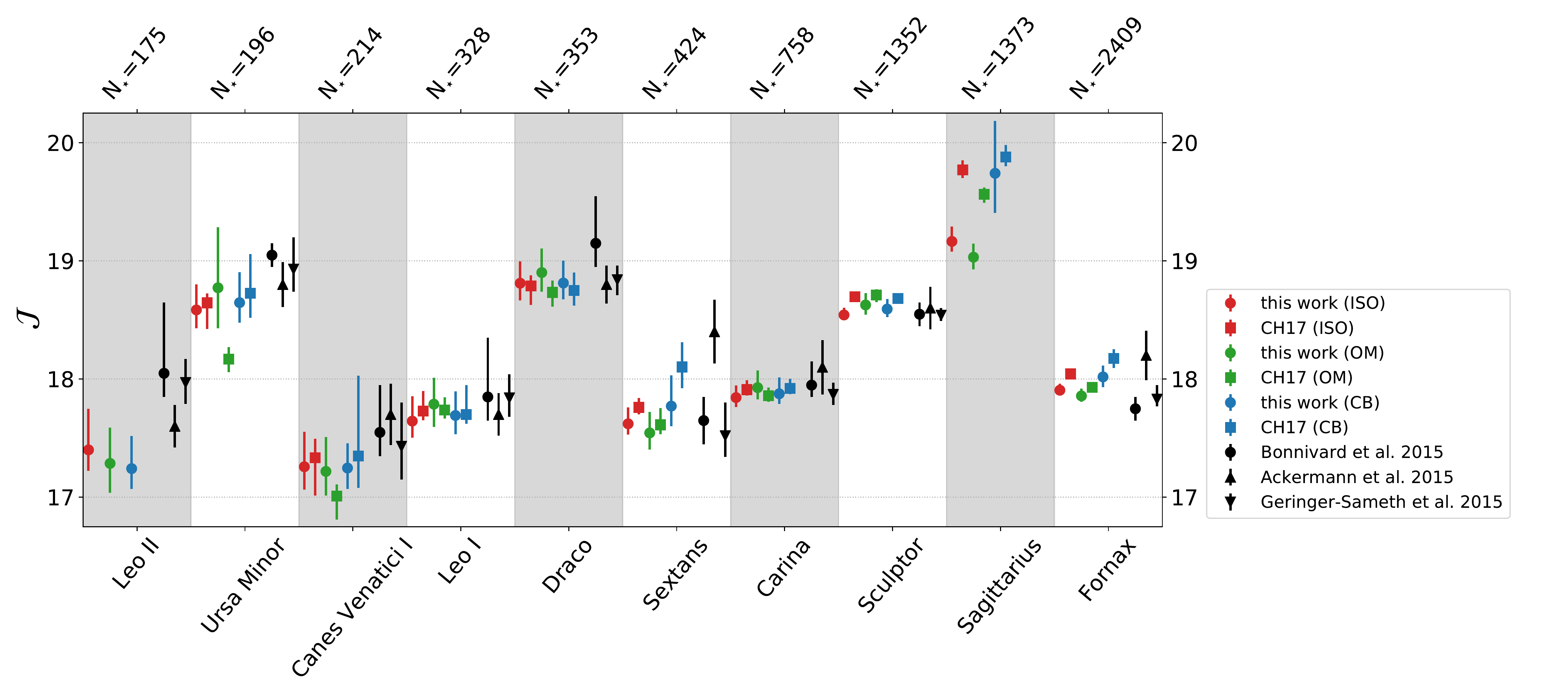}
\caption{Best-fit $\mathcal{J}$ values from the analysis of stellar 
kinematic data from 10 dSphs. The estimates are obtained from the 
approximation with Eq.~\ref{eq:linex} of the $\mathcal{J}$-envelope, 
constructed with the $J$-sampling scheme (circles). In all cases we 
assume a Plummer profile for the stellar brightness and a generalised 
NFW (Eq.~\ref{eq:genNFW}) for the DM profile. The red, green and blue 
symbols refer to the cases where we adopt the isotropic, OM and 
constant-$\beta$ velocity kernel functions, respectively.
For comparison, the analogous values released by CH17 are included 
(squares). The black points represent previous Bayesian-derived 
results used by \citet[upward triangles]{Ackermann:2015zua} 
and obtained by \citet[downward triangles]{Geringer-Sameth:2014yza} 
and \citet[circles]{Bonnivard:2015pia}.
The uncertainties on the estimates correspond to the 1$\sigma$ 
confidence intervals.}
\label{fig:Jall}
\end{figure*}

\begin{table}
\caption{Properties of the MW dSphs considered in this work. 
Values taken from \citet{Fermi-LAT:2016uux} and \citet{2012AJ....144....4M}.}
\setlength{\tabcolsep}{4pt}
\centering
\begin{tabular}{l r c c r c }
 Name & $l,b$ & Distance & $r_{1/2}$ & $M_{V}$  & $N_\star$ \\
 & (deg, deg) & (kpc) & (pc) & (mag) & \\
 \hline
 Canes Venatici I	& $74.31, 79.82$   & 218 & 441  & $-8.6$  & 214 \\
 Carina             & $260.11, -22.22$ & 105 & 205  & $-9.1$  & 758 \\
 Draco              & $86.37, 34.72$   & 76  & 184  & $-8.8$  & 353 \\
 Fornax             & $237.10, -65.65$ & 147 & 594  & $-13.4$ & 2409 \\
 Leo I              & $225.99, 49.11$  & 254 & 223  & $-12.0$ & 328 \\
 Leo II             & $220.17, 67.23$  & 233 & 164  & $-9.8$  & 175 \\
 Sagittarius        & $5.6, −14.2$     & 26  & 400  & $-13.5$ & 1373 \\
 Sculptor           & $287.53, -83.16$ & 86  & 233  & $-11.1$ & 1352 \\
 Sextans            & $243.50, 42.27$  & 86  & 561  & $-9.3$  & 424 \\
 Ursa Minor         & $104.97, 44.80$  & 76  & 120  & $-8.8$  & 196 \\
 \end{tabular}
\label{tab:dsphsprops}
\end{table}
Having explored the statistical properties of our frequentist $J$-sampling 
method, we proceed to applying it on data from real dSphs. To comply with 
the considerations presented above (Sec.~\ref{sec3}), we consider the dSphs with 
the most abundant kinematic sample available in the literature \citep{2012AJ....144....4M}, 
which results in ten galaxies with $N_\star \geq 100$. The systems and their 
properties are summarised in Table \ref{tab:dsphsprops}. In order to facilitate 
the comparison between our estimates and results in the literature, we always 
use $\Delta\Omega = 2.4 \times 10^{-4}$ sr, equivalent to $\theta_\textrm{max} = 0.5^\circ$. 
In accordance with the validation presented above, the profile likelihood 
curve for every dSph is built by approximating with Eq.~\ref{eq:linex} 
the output of the $J$-sampling approach (see Sec.~\ref{sec2}). The fit 
of each kinematic sample is repeated three times, implementing the ISO 
(Eq.~\ref{eq:keriso}), OM (Eq.~\ref{eq:kerom}) and CB (Eq.~\ref{eq:kerbeta}) 
kernel functions, respectively. 
We assume throughout a Plummer profile for the stellar density and 
a generalised NFW (Eq.~\ref{eq:genNFW}) for the DM distribution.
Therefore, the likelihood is sampled over the 5-(6-)dimensional 
parameter space $\mathcal{J}, r_0, a, b, c$ 
(plus $r_a$ or $\beta$).
The results are shown collectively in Fig.~\ref{fig:Jall}, which 
displays the best-fit $\mathcal{J}$ values, together with their 
uncertainty in the form of error bars reflecting the 1$\sigma$ 
confidence interval. 
The output of our method (circles) is compared with that of CH17 
(squares), when adopting the ISO (red),the OM (green) 
and the CB (blue) models. We also include recent Bayesian-derived 
estimates  used by \citet[upward black triangles]{Ackermann:2015zua} 
and those obtained by \citet[downward black triangles]{Geringer-Sameth:2014yza} 
and by \citet[black circles]{Bonnivard:2014kza}. 
Our results are also listed in Table \ref{tab:Jresults}, together 
with the best estimates of the parameter values of Eq.~\ref{eq:linex}. 
The full likelihoods corresponding to the data entering Fig.~\ref{fig:Jall} 
are shown in Fig.~\ref{fig:allikes} in the Appendix.

In most cases, the $\mathcal{J}_\textrm{MLE}$ values we obtain 
are consistent with the other sets of results, within quoted 
uncertainties. Although they use an analogous (frequentist) method, 
CH17 considered a less flexible model whereby they fixed the shape 
parameters in Eq.~\ref{eq:genNFW} to the NFW case. 
Despite assuming similar parameter ranges, the fitting methodology adopted 
by \citet{Geringer-Sameth:2014yza} differs substantially from our. 
In their work, these authors implemented \textit{MultiNest} 
\citep{Feroz:2013hea} to explore a 6-dimensional parameter space. Hence, 
the \citet{Geringer-Sameth:2014yza} estimates of the $\rho_\textrm{DM}$ 
parameters are influenced by the prior probability density. 
Although \citet{Ackermann:2015zua} examined a larger dimensionality 
parameters space, this does not match the one considered here. 
In particular, \citet{Ackermann:2015zua} don't fit the parameters of 
the profiles entering Eq.~\ref{eq:MLJeans}. Instead, those authors 
infer prior ranges on two characteristics of DM halos 
($v_\textrm{max}, r_\textrm{max}$) by assuming several (fixed) parametrisations 
of $\rho_\textrm{DM}$ (see \citealt{2015MNRAS.451.2524M} 
for details). Moreover, the same authors acknowledge the significant effect 
of prior choices on the marginalised posterior probability of $\mathcal{J}$ 
they derive (see Fig. 1 of \citealt{2015MNRAS.451.2524M}). 
\begin{table*}
\caption{Tabulated values of the $\mathcal{J}_\textrm{MLE}$ estimates entering  
Fig.~\ref{fig:Jall}. Columns 4,5,6 (8,9,10) [12,13,14] contain the best-fit 
parameters of the Eq.~\ref{eq:linex} approximating the $\mathcal{J}$-envelope, 
in turn obtained from the $J$-sampling scheme when assuming an ISO (OM) [CB]  
velocity anisotropy model.}
\centering
\setlength{\tabcolsep}{4pt}
\setlength\extrarowheight{4pt}
\begin{tabular}{| l c | c r r c | c r r c | c r r c |}
\hline
Dwarf	&	
$N_\star$ & 
$\mathcal{J}_\textrm{MLE}$ (ISO) & $p$ & $q$ & $r$ &
$\mathcal{J}_\textrm{MLE}$ (OM)  & $p$ & $q$ & $r$ &
$\mathcal{J}_\textrm{MLE}$ (CB)  & $p$ & $q$ & $r$ \\
\hline
Leo II             &  175 & ${17.40}^{+0.35}_{-0.18}$ & $  7.23$ & $  2.38$ & $ -0.09$ & ${17.29}^{+0.30}_{-0.25}$ & $  2.31$ & $  5.80$ & $ -0.45$ & ${17.24}^{+0.28}_{-0.17}$ & $  4.98$ & $  4.25$ & $ -1.02$ \\
Ursa Minor         &  196 & ${18.58}^{+0.21}_{-0.16}$ & $  4.66$ & $  6.47$ & $ -0.99$ & ${18.77}^{+0.51}_{-0.34}$ & $  2.96$ & $  2.00$ & $ -0.86$ & ${18.65}^{+0.26}_{-0.17}$ & $  4.06$ & $  5.57$ & $ -1.02$ \\
Canes Venatici I   &  214 & ${17.26}^{+0.30}_{-0.19}$ & $  5.51$ & $  3.33$ & $ -1.07$ & ${17.22}^{+0.29}_{-0.21}$ & $  4.77$ & $  3.66$ & $ -1.05$ & ${17.25}^{+0.21}_{-0.18}$ & $  5.18$ & $  5.51$ & $ -1.01$ \\
Leo I              &  328 & ${17.64}^{+0.21}_{-0.14}$ & $  5.32$ & $  6.45$ & $ -1.31$ & ${17.79}^{+0.22}_{-0.19}$ & $  2.82$ & $  8.44$ & $ -0.60$ & ${17.69}^{+0.20}_{-0.16}$ & $  4.19$ & $  7.40$ & $ -1.21$ \\
Draco              &  353 & ${18.81}^{+0.19}_{-0.15}$ & $  5.89$ & $  6.45$ & $ -1.02$ & ${18.90}^{+0.20}_{-0.16}$ & $  3.63$ & $  8.45$ & $ -1.11$ & ${18.81}^{+0.19}_{-0.14}$ & $  6.29$ & $  6.27$ & $ -1.07$ \\
Sextans            &  424 & ${17.62}^{+0.14}_{-0.09}$ & $  7.52$ & $ 10.50$ & $ -0.79$ & ${17.54}^{+0.18}_{-0.14}$ & $  5.26$ & $  7.75$ & $ -0.97$ & ${17.77}^{+0.26}_{-0.17}$ & $  6.98$ & $  3.44$ & $ -1.08$ \\
Carina             &  758 & ${17.84}^{+0.10}_{-0.08}$ & $ 11.82$ & $ 10.74$ & $ -0.77$ & ${17.93}^{+0.15}_{-0.10}$ & $  6.62$ & $ 10.29$ & $ -1.20$ & ${17.88}^{+0.14}_{-0.09}$ & $ 11.57$ & $  7.41$ & $ -1.02$ \\
Sculptor           & 1352 & ${18.54}^{+0.06}_{-0.04}$ & $ 16.74$ & $ 22.11$ & $ -1.03$ & ${18.63}^{+0.10}_{-0.08}$ & $  9.48$ & $ 13.21$ & $ -1.82$ & ${18.59}^{+0.08}_{-0.07}$ & $ 13.61$ & $ 13.48$ & $ -1.23$ \\
Sagittarius        & 1373 & ${19.16}^{+0.13}_{-0.09}$ & $  9.78$ & $  9.50$ & $ -0.84$ & ${19.03}^{+0.11}_{-0.10}$ & $  7.01$ & $ 12.39$ & $ -1.29$ & ${19.74}^{+0.44}_{-0.34}$ & $  1.90$ & $  3.58$ & $ -0.87$ \\
Fornax             & 2409 & ${17.91}^{+0.05}_{-0.04}$ & $ 22.70$ & $ 19.47$ & $ -0.77$ & ${17.86}^{+0.06}_{-0.05}$ & $ 14.67$ & $ 22.05$ & $ -1.22$ & ${18.02}^{+0.10}_{-0.09}$ & $  9.36$ & $ 13.08$ & $ -1.67$ \\
\hline
\end{tabular}
\label{tab:Jresults}
\end{table*}

We stress that our statistical approach (profile likelihood for frequentist 
$J$-factors) is not dependent on the model assumptions, and thus is easily 
extendable to other systems, such as faint dSphs, once more kinematic data 
is provided. Moreover, this procedure can be applied to all galaxies where 
spectroscopic observations are available, \textit{e.g.} field galaxies or their 
satellite galaxies. However, in the former case the modelling might change, 
with the spherical Jeans equation likely not being warranted any more. 
Additionally, the contribution of stars to the gravitational potential would 
need to be incorporated in the model. In this situation, the use of the 
Jeans formalism should possibly be abandoned, in favour of a physically motivated 
velocity distribution function. The derivation of this quantity and its 
implementation in the frequentist schemes presented here is currently under 
development and will be presented in a future publication.

\section{Consistent Dark Matter annihilation cross-section limits}
\label{sec5}
In this section we implement the profile likelihoods of $\mathcal{J}$ with 
the $\gamma$-ray data, for the latter using the Fermi-LAT observations, 
producing consistent DM annihilation cross-section limits. 
We use all curves derived in Sec.~\ref{sec4}, with the exception of Sagittarius. 
This dSph is the closest to the MW (Table \ref{tab:dsphsprops}) 
and is known to be experiencing strong tidal disruption by the MW potential 
\citep{Johnston:1995vd}. Similarly, most previous analyses of $\gamma$-ray 
data neglect this system when searching for possible signals of annihilating 
DM in dSphs of the Local Group 
\citep{Ackermann:2015zua, 2015ApJ...809L...4D, Fermi-LAT:2016uux}.

\begin{figure}
\centering
\includegraphics[width=.5\textwidth, keepaspectratio]{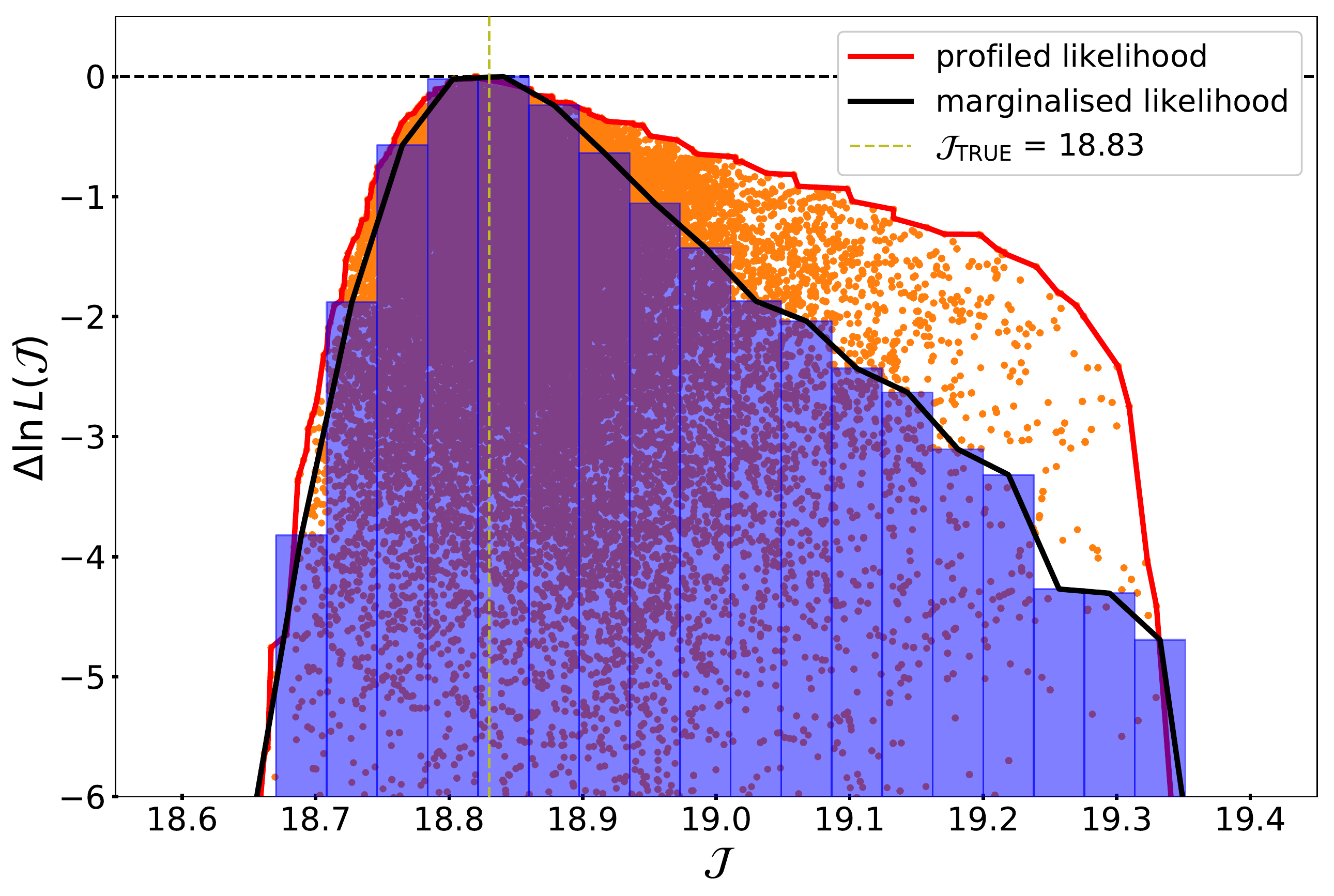}
\caption{Comparison between profiling and marginalising the likelihood 
over the nuisance parameters. The orange points indicate the likelihood 
evaluations over the allowed parameter space. The marginalisation 
of these points gives the blue histogram. The different width of the 
$J$-envelope (red curve) and the marginalised likelihood (black curve) at 
large $\mathcal{J}$ displays the effect on the uncertainties on the 
parameter yielded by the two statistical approaches. The stellar kinematic 
data used when sampling the likelihood belongs to one of the PE analysed in 
Sec.~\ref{sec3}. The vertical dashed line indicates the true value of 
$\mathcal{J}$ for the underlying GAIASIM model.}
\label{fig:marg_prof}
\end{figure}

It is customary in astronomy to compare an observed photon flux 
with the expectations via a Poisson likelihood. Clearly, in doing so, 
the contamination from concomitant spurious sources, such as the galactic 
diffuse emission and point sources, must be adequately taken into account. 
Performing this subtraction for the LAT data entails an analysis 
using the \textit{Fermi Science Tools} 
\footnote{\url{https://fermi.gsfc.nasa.gov/ssc/data/analysis/software/}}. 
Conveniently, \citet{Ackermann:2015zua} have published their results in 
the form of bin-by-bin likelihood tables (see \citealt{Ackermann:2015zua} and 
reference therein for more details). Inference of DM properties proceeds 
then via the optimisation of the following function \citep{Ackermann:2015zua}
\begin{eqnarray}
\mathcal{L}_\textrm{d}(\bm{\mu},\bm{\Xi}=
(\bm{\alpha},J_\textrm{d})|\mathcal{D^\textrm{d}}=
(\mathcal{D}^\textrm{d}_\gamma,\mathcal{D}^\textrm{d}_\star))= \nonumber \\
-\ln P_\textrm{d}(\bm{\mu},\bm{\alpha}, 
J_\textrm{d}|\mathcal{D}^\textrm{d}_\gamma) \, +  
\mathcal{L}_J(J_\textrm{d}|\mathcal{D}^\textrm{d}_\star) \, .
\label{eq:globalL}
\end{eqnarray}
In Eq.~\ref{eq:globalL}, the vector $\bm{\mu}$ represents the parameters 
of interest $(m_\textrm{DM},\left<\sigma v\right>)$, while $\bm{\Xi}$ 
contains the nuisance parameters involved in the Poisson likelihood 
($P$) optimisation, $\bm{\alpha}$, and the $J$-factor. 
$\mathcal{D}^\textrm{d}$ comprises the photon data, $\mathcal{D}^\textrm{d}_\gamma$, and the stellar kinematic sample, 
$\mathcal{D}^\textrm{d}_\star$. The index 'd', appearing in most terms of 
Eq.~\ref{eq:globalL}, indicates that these quantities refer to a specific 
dSph. Since the properties of a given DM candidate are assumed to be 
invariant across different targets, it is possible to combine the likelihood 
for each dSph  (Eq.~\ref{eq:globalL}) into a unique object. Optimising the 
following joint likelihood \citep{Ackermann:2011wa} 
\begin{equation}
\widetilde{\mathcal{L}}(\bm{\mu}) = \prod^{N_\textrm{dwarfs}}_\textrm{d=1} 
\mathcal{L}_\textrm{d}(\bm{\mu}, \bm{\Xi} | \mathcal{D}^\textrm{d}) \, ,
\label{eq:combL}
\end{equation}
thus increases the sensitivity over individual targets. From 
Eq.~\ref{eq:globalL} we see that $\mathcal{L}_J(J)$ plays a decisive 
role in the determination of the DM particle properties. Optimising 
this formula with respect to $J$ allows the propagation of astrophysical 
uncertainties when constraining $\left<\sigma v\right>$. The importance of 
using a frequentist-derived $J$-factor likelihood in Eq.~\ref{eq:globalL} is 
displayed in Fig.~\ref{fig:marg_prof}. This figure compares the $J$-envelope 
(red curve) of the likelihood evaluations (orange dots) with the 
marginalisation of the same over all nuisance parameters (blue histogram). 
We note how both the profiled (red curve) and marginalised (black curve) 
likelihoods recover well the true $J$-factor. However, the marginalisation 
process produces a narrower curve than the former, especially at large 
$\mathcal{J}$ values. This feature implies that uncertainties obtained 
from the marginalised posterior are artificially smaller than what indicated 
by the profile likelihood.

\begin{figure}
\centering
\includegraphics[width=.5\textwidth, keepaspectratio]{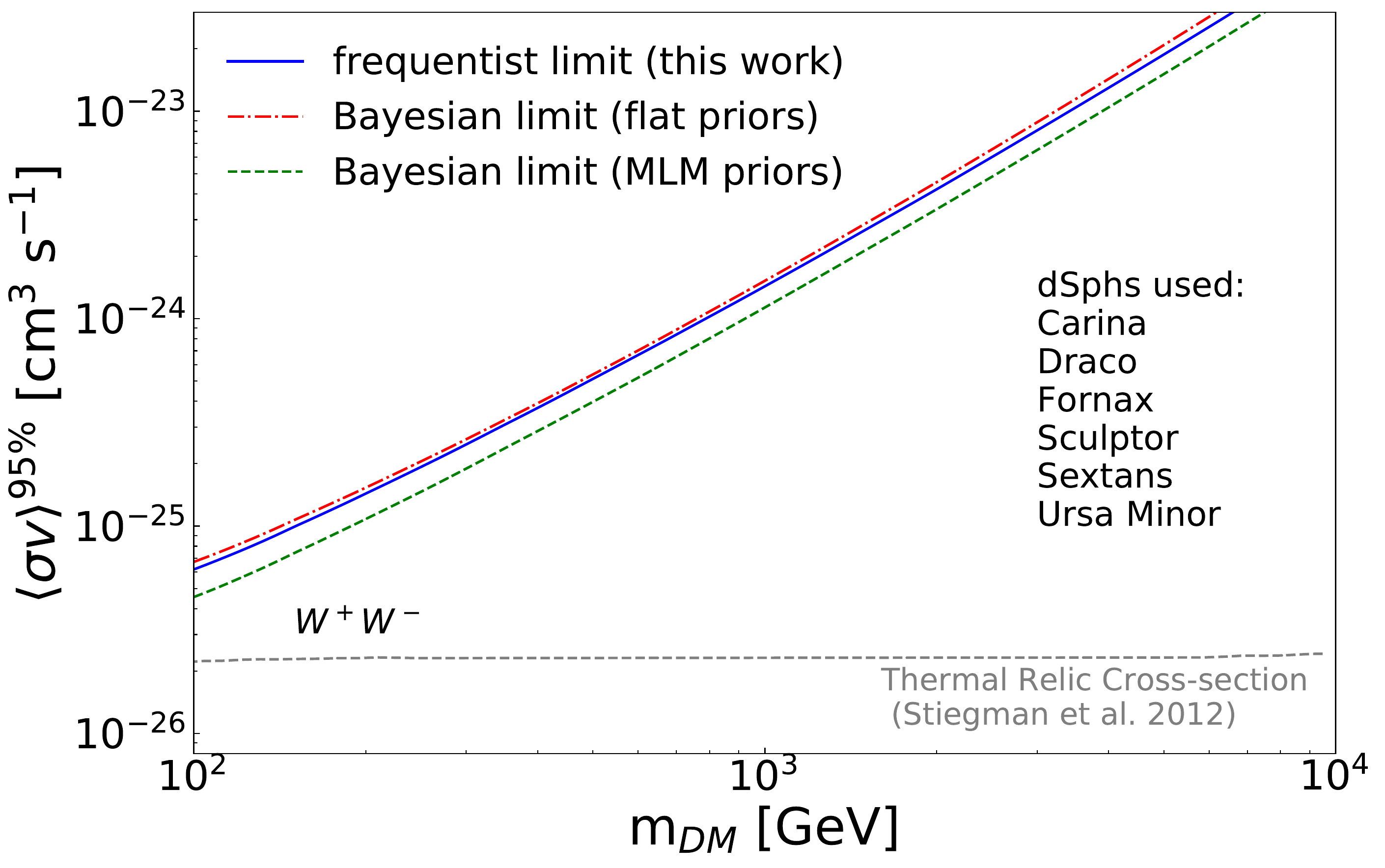}
\caption{DM annihilation cross-section 95 per cent upper limits from a combined 
analysis of six dSphs (see figure) using $J$-factor likelihoods derived with 
different statistical assumptions. The blue solid curve represents the new, 
consistent UL determined by implementing the $\mathcal{L}(\mathcal{J})$ 
curves derived with the frequentist method presented in this work 
(Sec.~\ref{sec2}). For comparison, the dot-dashed red (dashed green) line 
constitutes the analogous result calculated with the $J$-factor likelihood 
parameterisation adopted by \citealt{Ackermann:2011wa} (\citealt{Ackermann:2015zua}). 
The curves are obtained assuming that all DM annihilates through the $W^+W^-$ 
channel. The dotted grey curve is the thermal relic cross section derived 
in \citet{Steigman:2012nb}.}
\label{fig:uplimPcfr}
\end{figure}

Setting limits on the DM annihilation cross-section with $\widetilde{\mathcal{L}}$ entails, for a fixed $m_\textrm{DM}$, finding the 
largest value of $\left<\sigma v\right>$ for which $\widetilde{\mathcal{L}}$ 
increases by $2.71/2$ from its minimum. Repeating this process for a range 
of a masses yields a curve in the $(m_\textrm{DM},\left<\sigma v\right>)$ 
plane, which provides a 95\% confidence level upper limit (UL) on 
$\left<\sigma v\right>$. The product of this operation is reported in 
Fig.~\ref{fig:uplimPcfr}, where our result (blue solid curve) is displayed 
together with the analogous one calculated implementing the $J$ likelihood 
parameterisations proposed by \citet[red dot-dashed curve]{Ackermann:2011wa} 
and \citet[green dashed curve]{Ackermann:2015zua}. We reiterate that the $J$ 
likelihood curves adopted in these works were obtained via Bayesian analyses 
of stellar kinematic data, implementing flat priors in the former and the MLM 
priors of \citet{2015MNRAS.451.2524M} in the latter. Therefore, 
Fig.~\ref{fig:uplimPcfr} provides a direct comparison of the UL on 
$\left<\sigma v\right>$ stemming from the different statistical configurations. 
The dSphs sample chosen for the three joint likelihood analyses is dictated 
by the set of targets considered in \citet{Ackermann:2011wa} and 
\citet{Ackermann:2015zua}: for consistency, we select the subset of dSphs 
common to both publications. We are, thus, left with the following systems: 
Carina, Draco, Fornax, Sculptor, Sextans and Ursa Minor. For illustrative 
reasons, we assume that all DM annihilates into $W^+W^-$ pairs. Unsurprisingly, 
the UL derived from the flat-prior $J$ likelihoods is very similar to that 
calculated with our profile likelihood curves of $J$. This resemblance follows 
the observation that our likelihood sampling expedient is, essentially, an 
ordinary MCMC where the priors are deprived of their numerical influence on 
the likelihood. Moreover, the targets sample considered contains the brightest 
known dSphs of the MW and Bayesian statistics progressively becomes less 
sensitive to priors as the number of observations grows. Following the previous 
point, the stronger constraining power of the MLM UL (shown in green) is 
principally determined by the action of the MLM priors in the (Bayesian) 
analysis of the kinematic data by \citet{2015MNRAS.451.2524M}. Importantly, 
Fig.~\ref{fig:uplimPcfr} elucidates the advantage of performing a frequentist 
analysis of stellar data over the standard Bayesian framework: removing the 
effect of priors, we eliminate their influence on the DM annihilation 
cross-section ULs, associated with the arbitrariness of their selection. 

\begin{figure}
\centering
\includegraphics[width=.5\textwidth, keepaspectratio]{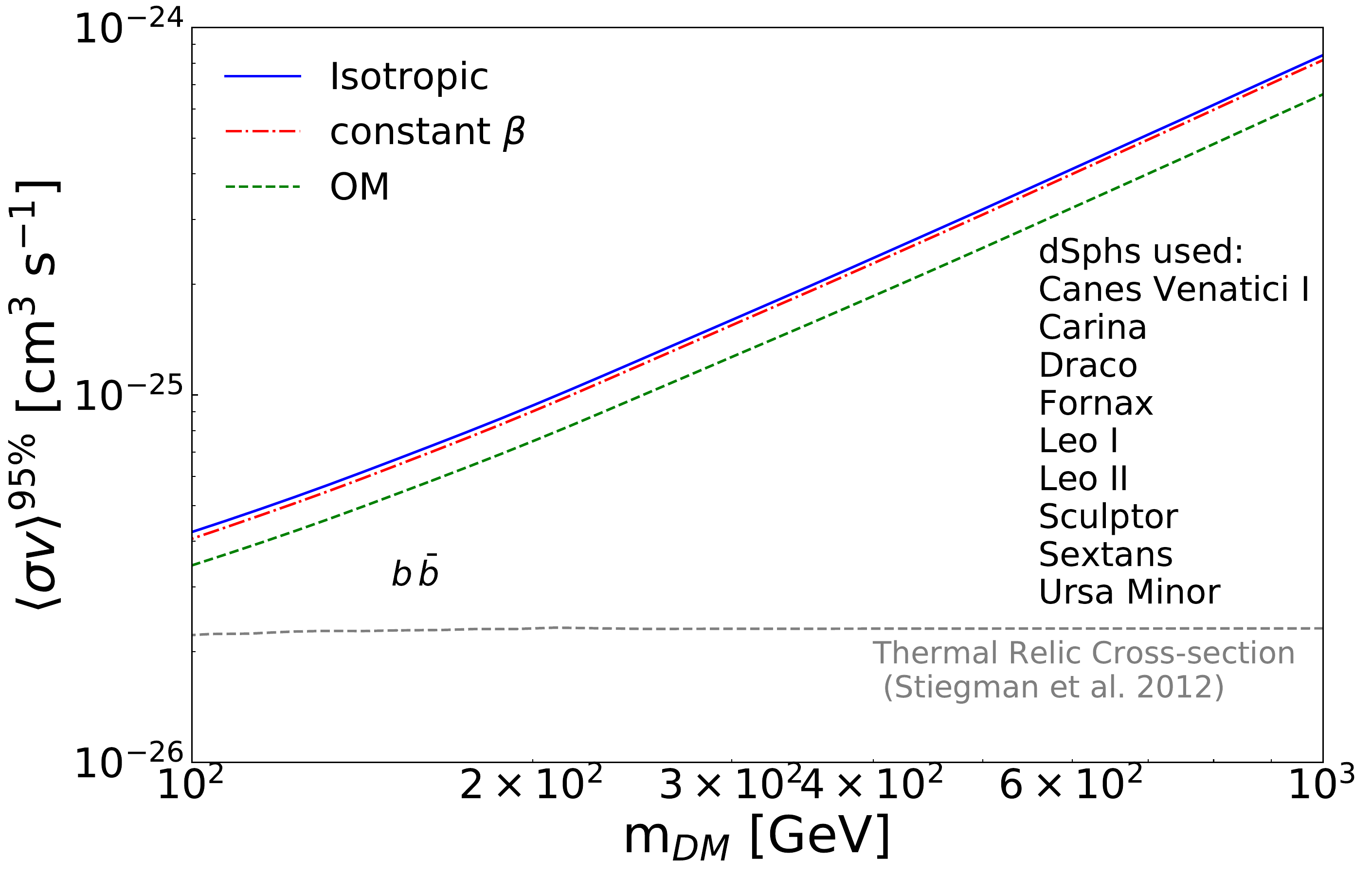}
\caption{Same as Fig.~\ref{fig:uplimPcfr}, but obtained considering 
nine dSphs (see figure) and using $J$-factor likelihoods built by implementing 
different models of the stellar velocity anisotropy. Specifically, we consider 
the case of isotropic velocities (solid blue line), constant degree of 
anisotropy $\beta$ (dot-dashed red line) and the OM profile 
(dashed green line). The curves are derived assuming that all DM 
annihilates into $b\bar{b}$ quarks. The dotted grey curve represents the 
thermal relic cross section derived in \citet{Steigman:2012nb}.}
\label{fig:uplimJcfr}
\end{figure}

The frequentist UL shown in Fig.~\ref{fig:uplimPcfr} is calculated adopting 
the likelihood curves of $J$ derived in the isotropic stellar velocities 
assumption. This choice guarantees the modelling consistency underlying 
the $J$ likelihood component of the UL determination entering the figure. The 
effect of implementing different velocity anisotropy models on the DM 
annihilation ULs is portrayed in Fig.~\ref{fig:uplimJcfr}. 
In producing this plot, we consider a broader sample of dSphs, consisting 
of all systems listed in Table \ref{tab:Jresults} except Sagittarius, 
due to the arguments presented at the beginning of this section. Moreover, 
we assume that all DM annihilates into $b\bar{b}$ quark pairs, when 
the stellar velocities are isotropically oriented (Eq.~\ref{eq:keriso}, 
solid blue line), when they have a constant degree of anisotropy throughout 
the dSphs (Eq.~\ref{eq:kerom}, dot-dashed red line) or have an OM radial 
velocity anisotropy profile (Eq.~\ref{eq:kerbeta}, dashed green line). The 
similarity of these curves is driven by the affinity of the corresponding 
$J$ likelihood curves for most dSphs considered -- shown in Fig.~\ref{fig:allikes} 
in the Appendix -- and the proximity of their $\mathcal{J}_\textrm{MLE}$ values, 
especially.

\section{Conclusion}
The high energy radiation from dSphs may be the key to DM indirect 
identification. Undesirably, the inaccessibility of the spatial distribution 
of DM within dSphs undermines current searches by introducing a major 
source of systematic uncertainty. Some of this indeterminacy has been 
typically tamed by means of Bayesian techniques. Inevitably, though, 
the presence of priors in this kind of analysis entails potential bias 
on the estimates, in this case the MLE $J$-factor. Moreover, 
the optimisation of Eq.~\ref{eq:globalL} with respect to $J$ implies 
the propagation of the effect of priors to the final result: the 
DM annihilation $\left<\sigma v\right>^{95\%}$ UL. Since the 
Poisson likelihood in Eq.~\ref{eq:globalL} is usually obtained in a 
frequentist manner, the use of such Bayesian-derived $J$-factor likelihoods 
entails an inconsistency in the statistical approach employed to derive 
the $\left<\sigma v\right>^{95\%}$ UL.

Adopting the reformulation of $J$ proposed by CH17 (Eq.~\ref{eq:rho0}), 
with an expedient for using the MCMC within a frequentist framework, 
we devise a scheme for fitting a generalised Jeans equation to the stellar 
kinematics of a dSph, removing the need for priors. The validation of our 
method on simulations (Sec.~\ref{sec3}) indicates that this frequentist 
approach possesses satisfactory statistical properties, when analysing 
sufficiently large kinematic samples (i.e. for $N_\star \geq 100$).
Following this prescription, we obtain new, prior-free profile likelihoods 
for the $J$-factor of 10 bright dSphs. The MLE values of $J$ we derive 
are broadly consistent with previous results found in the literature. 

Implementing the new likelihood curves in the optimisation of Eq.~\ref{eq:combL}, 
provides the first statistically-consistent (fully-frequentist) ULs on 
the DM annihilation cross-section. From multiple joint-likelihood analyses 
of six dSphs, we compare the new constraints with those obtained implementing 
the (Bayesian-derived) parameterisations of the likelihood of $J$ proposed by 
\citet[using flat priors]{Ackermann:2011wa} and 
\citet[involving MLM priors]{Ackermann:2015zua}. An advantage of our method 
is the removal of the potential arbitrariness related to the choice of priors. 
Additionally, we present $\left<\sigma v\right>^{95\%}$ ULs on the DM 
annihilation associated with different assumptions on the velocity anisotropy 
of the stellar population of dSphs. The similarity of the constraints (and 
$J$-factors -- see Fig.~\ref{fig:allikes}) for the different models considered 
in this work (Eq.~\ref{eq:ker}) could be due to the \textit{mass-anisotropy} 
degeneracy afflicting the mass determination in dSphs \citep{Wolf:2009tu}. 

A possible venue of improvement of this method concerns the use of a 
Gaussian likelihood for the los velocities of stars (Eq.~\ref{eq:lnlike}). 
This \textit{ad-hoc} assumption could be replaced, for example, by adopting 
3-dimensional Gaussian stellar velocities, as done in the \textsc{MAMPOSSt} 
routine \citep{2013MNRAS.429.3079M}. Alternatively, the dynamics of stars in 
dSphs can be modelled via action-angle 
variables, as done in the AGAMA code \citep{2019MNRAS.482.1525V}; 
for a recent review on the topic, see \citet{2016MNRAS.457.2107S}. The solution 
that we intend to pursue in the future entails the derivation of the physical 
velocity distribution function of the system from the Eddington inversion formula 
\citep{2008gady.book.....B}. We note that this calculation can be performed 
only for the stellar velocity anisotropy models considered here 
\citep{2008gady.book.....B}. Recent applications of this technique to the stellar 
population of the MW have proven to reproduce quite accurately the observations 
\citep{2014MNRAS.445.3133P,2014IAUS..298..117B,2014ApJ...793...51S}. 
Moreover, the use of the Gaussian approximation may be the culprit for the 
sub-optimal statistical properties of the profile likelihood resulting 
from the fitting schemes presented in Section \ref{sec2}. We also note 
that the performance of the method can be improved by implementing 
an exponential cut-off in the DM distribution, as done in previous 
works (see \citealt{Lokas:2004sw} and references therein). However, 
we take the effect of this modification -- which mimics the tidal 
stripping by the potential of the Milky Way \citep{Kazantzidis:2003im} -- 
to be incorporated into the approximation with Eq.~\ref{eq:linex} 
(see Sec.~\ref{sec2}). Generally, our method depends on the availability 
of abundant kinematic samples, ideally extending to large projected 
radial distances. This assertion implies that the analysis of \textit{ultra-faint} 
dSphs will inevitably necessitate more spectroscopic observations. 
Extensive stellar data for these systems will allow to implement a 
joint-likelihood analysis for a broader ensemble 
of targets, thereby setting new, data-driven ULs on $\left<\sigma v\right>$. 
Additionally, we will be able to compare the frequentist constraints 
with the Bayesian ones (for example of \citealt{Ackermann:2015zua}) 
and test the claimed detection of $\gamma$ rays in Ret II dSph 
\citep{Geringer-Sameth:2015lua}. We, therefore, auspicate that future surveys 
will perform accurate and extended spectroscopic observations of ultra-faint 
and newly discovered dSphs, for example those recently detected by the 
DES observatory \citep{Bechtol:2015cbp,Drlica-Wagner:2015ufc}.

\section*{Acknowledgements}
AC is thankful to Knut Dundas Mor\aa\, and Stephan Zimmer for useful 
discussions. The work of LES is supported by DOE Grant de-sc0010813. 
JC is supported by grants of the Knut and Alice Wallenberg Foundation 
and the Swedish Research Council. 

\bibliographystyle{mnras}
\bibliography{paper}
\bsp	

\appendix
\label{appendix}
\section{Profile likelihood curves}
\begin{figure*}
\includegraphics[width=0.9\textwidth,keepaspectratio]{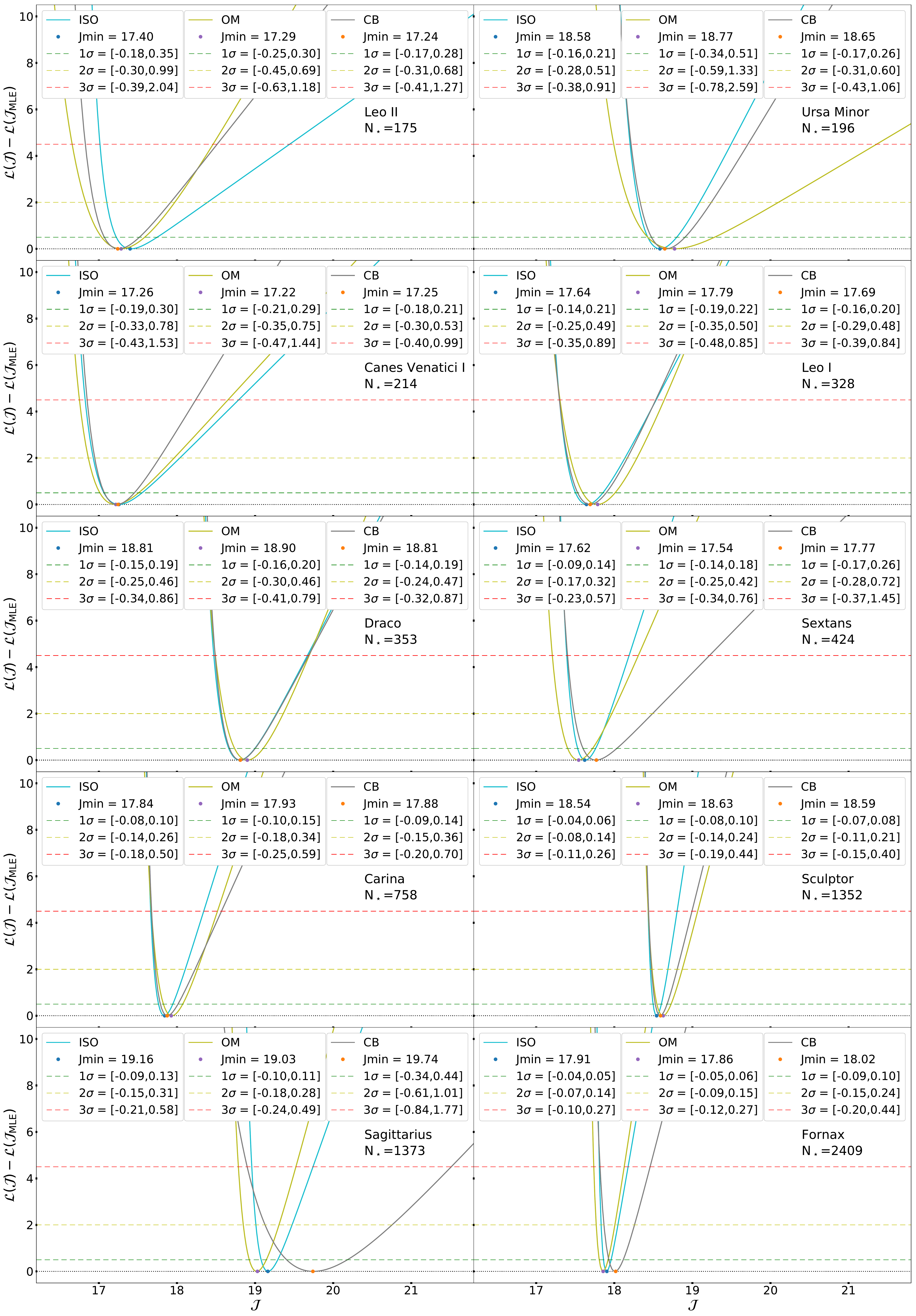}
\caption{Full likelihoods for the ten dSphs considered in this 
work. Each panel refers to a given system, as indicated in the images. The 
curves shown represent the approximation with Eq.~\ref{eq:linex} of the 
likelihood resulting from the $J$-sampling scheme when implementing the 
isotropic (cyan line), OM (olive line) and constant-$\beta$ (grey line) 
velocity anisotropy models. The best-fit $\mathcal{J}$ values and the 
width of the 1,2,3$\sigma$ intervals are also indicated in each panel, 
for each curve.}
\label{fig:allikes}
\end{figure*}

\label{lastpage}
\end{document}